\documentclass[conference]{IEEEtran}

\usepackage{cite}
\usepackage{fancyheadings}
\usepackage{graphicx}
\usepackage{amssymb}
\usepackage{amsmath}
\usepackage{latexsym}
\usepackage{bm}
\usepackage{multirow}
\usepackage{color}
\usepackage{array}
\usepackage{slashbox}
\newcolumntype{L}[1]{>{\raggedright\let\newline\\\arraybackslash\hspace{0pt}}m{#1}}
\newcolumntype{C}[1]{>{\centering\let\newline\\\arraybackslash\hspace{0pt}}m{#1}}
\newcolumntype{R}[1]{>{\raggedleft\let\newline\\\arraybackslash\hspace{0pt}}m{#1}}

\newtheorem{theorem}{Theorem}

\newtheorem{lemma}{Lemma}
\newtheorem{definition}{Definition}

\newtheorem{example}{Example}

\newcommand{\triequ}{\stackrel{{\scriptscriptstyle{\bigtriangleup}}}{=}}
\newcommand{\myQED}{\mbox{}\hfill{$\Box$}}

\begin{document}

\title{Construction of Unrestricted-Rate\\ Parallel Random Input-Output Code}

\author{\IEEEauthorblockN{Shan Lu, Horoshi Kamabe}
\IEEEauthorblockA{Department of Electrical, Electronic,\\ and Computer Engineering,\\
Gifu University, \\
1-1, Yanagido, Gifu, 501-1193 Japan\\
Email: shan.lu.jp@ieee.org}
\and
\IEEEauthorblockN{Jun Cheng}
\IEEEauthorblockA{Department of Intelligent Information\\
Engineering and Sciences, \\
 Doshisha University,\\
 Kyotanabe, Kyoto, 610-0321 Japan\\
 Email: jcheng@ieee.org}
 \and
 \IEEEauthorblockN{Akira Yamawaki}
\IEEEauthorblockA{Department of Electrical, Electronic,\\ and Computer Engineering,\\
Gifu University, \\
1-1, Yanagido, Gifu, 501-1193 Japan\\
Email: yamawaki@kmb.info.gifu-u.ac.jp}
}

\maketitle

\begin{abstract}
A coding scheme for two-page unrestricted-rate P-RIO code that each page may have different code rates is proposed. In the second page, the code for each messages consists of two complementary codewords with code length $n$. There are a total of $2^{n-1}$ codes which are disjoint to guarantees uniquely-decodable for $2^{n-1}$ messages. In the first page, the code for each message consists of all weight-$u$ vectors with their non-zero elements restricted to $(2u-1)$ same positions, where non-negative integer $u$ is less than or equal to half of code length. Finding codes to be disjoint in first page is equivalent to construction of constant-weight codes, and the numbers of disjoint codes are the best-known numbers of codewords in constant-weight codes. Our coding scheme is constructive, and the code length is arbitrary. The sum rates of our proposed codes are higher than those of previous work.
\end{abstract}

\section{Introduction}
Flash memories are the prevalent type of non-volatile memory (NVM) in use today which are intended for SSD and mobile applications. Flash memories are comprised of blocks of cells. The cells can have binary values or multiple levels. Multilevel flash memories can store multiple bits in a cell. Conventionally, in multilevel flash memory, in order to read a single logical page, more than a single read threshold, on average, is required.

To increase input/output (I/O) performance for multilevel flash memories. A random input-output (RIO) code \cite{Sharon2013} is proposed. It is a coding scheme that permits writing $q-1$ pages in $q$ levels and reading one page of data from multi-level flash memories only uses one single read threshold.
A one-to-one correspondence between RIO codes \cite{Sharon2013} and the well studied WOM codes \cite{Rivest1982} is shown. However, in WOM codes, the encoder sets the cell state values based on the current memory state and the received message on each write, thus, the message are stored sequentially and are not all known in advance while encoding. Moreover, it is difficult to use difference code rate for each writing in WOM codes.

In RIO codes, all the messages of each level can be known in advance, it is unnecessary to store messages sequentially and possible to control difference code rate for each level. Therefore, a parallel RIO (P-RIO) code \cite{Yaakobi2016} is proposed that the encoding of each level is done in parallel while reading one page of data using a single read threshold. Higher sum-rates of P-RIO codes are achieved \cite{Yaakobi2016} than that of  RIO codes \cite{Sharon2013}. In \cite{Yaakobi2016}, only fixed-rate P-RIO code is considered that the rates of each page are same. The complexity of research algorithm in \cite{Yaakobi2016} increases exponentially and the search space quickly becomes memory and computationally intensive as code length increases.

In this paper, we propose a coding scheme for two-page unrestricted-rate P-RIO code that each page may have different code rates. In the second page, the code for each messages consists of two complementary codewords with code length $n$. There are a total of $2^{n-1}$ codes which are disjoint to guarantees uniquely-decodable for $2^{n-1}$ messages. In the first page, the code for each message consists of all weight-$u$ vectors with their non-zero elements restricted to $(2u-1)$ same positions, where non-negative integer $u$ is less than or equal to half of code length. Finding codes to be disjoint in first page is equivalent to construction of constant-weight codes, and the numbers of disjoint codes are the best-known numbers of codewords in constant-weight codes. Our coding scheme is constructive, and the code length is arbitrary. The sum rates of our proposed codes are higher than those of conventional fixed-rate P-RIO codes in \cite{Yaakobi2016}.

\section{Preliminary}
\subsection{Unrestricted-Rate P-RIO Code}
Assume that in the flash memory, the cells have $q$ levels and all cells are in level zero initially. It is only possible to increase the level of each cell. We denote by $[q] = \{0, 1, \dots, q-1\}$. For $n$ cells, a vector $\bm{x}=(x_0, x_1, \dots, x_{n-1}) \in [q]^{n}$ will be called a cell state vector.

P-RIO code is a coding scheme that encoding of each page (level) in parallel while reading one page of data using a single read threshold. In a $q$-level flash memory, there are $(q-1)$'s pages. Let $M_i$, $i=1,2, \dots q-1$, be the number of messages for $i$th page.

We expand the definition of P-RIO codes in \cite{Yaakobi2016} to $(n; M_1, M_2, \dots, M_{q-1})$-P-RIO code as follows.
\begin{definition} [\cite{Yaakobi2016}]
An $(n; M_1, M_2, \dots, M_{q-1})$-P-RIO code is a code with an encoding scheme comprising of $n$ cells with $q$ levels and is defined by encoding map $$\xi: [M_1] \times [M_2] \times  \dots \times [M_{q-1}]  \rightarrow [q]^n$$ and decoding maps $\mathfrak{D}_{i}: [q]^n \rightarrow [M_i]$, for $i = 1,2, \dots, q-1$, where messages of each page $(m_1,m_2, \dots, m_{q-1}) \in [M_1] \times [M_2] \times  \dots \times [M_{q-1}] $.   \myQED
\end{definition}

The sum rate of $(n; M_1, M_2, \dots, M_{q-1})$-P-RIO code is $$R_{\text{sum}} = \frac{\sum\limits _{i=1}^{q-1}\log_{2}{M_i}}{n}.$$
Note that  the code rates of each page in $(n; M_1, M_2, \dots, M_{q-1})$-P-RIO codes may be unequal, that is $M_i \not = M_j$ for $i \not=j$. In \cite{Yaakobi2016}, only the P-RIO codes with same code rate of each page are considered.

The decoding maps $\mathfrak{D}_{i}: [q]^n \rightarrow [M_i]$ is first to read one page of data using a single read threshold, and then decode the data to corresponding message.  Let $r$ be a threshold level between pages $r-1$ and $r$, $r=1,2, \dots, q-1$, called $r$th threshold. The data of reading the $r$th threshold from the cell state $\bm{x}=(x_0, x_1, \dots, x_{n-1}) \in [q]^{n}$ is denoted as $\bm {d}_{r}(\bm{x}) = (d_0, d_1, \dots, d_{n-1}) \in \{0,1\}^{n}$ where $$d_r(x_i) =\left\{\begin{array}{ll}
0, &  x_i <r   \\
1, &  x_i \geq r
\end{array}\right.$$
Vector $\bm {d}_{r}(\bm{x})$ gives the $(q-r)$th level's message.

\subsection{$(n; M_1, M_2)$-P-RIO Code}
Now we consider $q=3$ flash memory, that is to say, there are two pages on each cell.

Let $\bm{a}=(a_0,a_1, \dots, a_{n-1})$ and $\bm{b}=(b_0,b_1, \dots, b_{n-1})$ be two $n$-vectors. We say that vector $\bm{a}$ is included in vector $\bm{b}$, denoted by $\bm{a} \leq \bm{b}$,  if and only if $a_i \leq b_i$ for $i=0,1, \dots, n-1$.

Given two sets $A, B \subset \{0,1\}^{n}$, if there exists at least one vector $\bm{a} \in A$ and at least one vector $\bm{b}\in B$ such that $\bm{a} \leq \bm{b} $, we say that set $A$ is included in set $B$, denoted by $A \leq B$.

Let set $ \textsf{B}=\{B_0, B_1, \dots, B_{M-1}\}$ with $B_{j} \subseteq \{0,1\}^{n}$, $0 \leq j \leq M-1$. we say that $A$ is included in $\textsf{B}$, denoted by $A \leq \textsf{B}$, if $A \leq B_j$ for all $j=0,1,\dots, 2^{n-1}$.

The following theorem is given in \cite{Yaakobi2016} with $M_1=M_2$.
\begin{theorem}[\cite{Yaakobi2016}] \label{P-RIO}
An $(n; M_1, M_2)$ P-RIO code exists if and only if two sets $\textsf{A} =\{A_0, A_1, \dots, A_{M_1-1}\}$ and $\textsf{B}=\{B_0, B_1, \dots, B_{M_2-1}\}$ with $A_{i}, B_{j} \subseteq \{0,1\}^{n}$, $0 \leq i \leq M_{1}-1$, $0 \leq j \leq M_{2}-1$ that satisfy the following conditions:
\begin{enumerate}
 \item  $A_{i}\cap A_{i'} = \O$ and $B_{j} \cap B_{j'} = \O$ for all $0 \leq i, i' \leq M_{1}-1$, $0 \leq j ,j'\leq M_{2}-1$, $i \not= i', j\not=j'$.
 \item For any $i$, $i=0,1,\dots, M_1-1$,  $A_i \leq \textsf{B}$. \myQED
\end{enumerate}
\end{theorem}

From Theorem \ref{P-RIO}, we have that set $\textsf{A}$ is the code of first page, subset $A_i$ is the constituent code for the $i$th message. Similarly, $\textsf{B}$ is the code of second page, subset $B_j$ is the constituent code for the $j$th message. The first condition in Theorem \ref{P-RIO} guarantees the uniquely-decodable for each page's messages. The second condition guarantees reading one page of data using a single read threshold.

\begin{example}
Table~\ref{TPRIO3} gives an example of (3;5,4)  P-RIO code. The columns correspond to the symbol value of the first page. The sets of the code are $\textsf{A}=\{\{000\},\{001\},\{010\},\{100\},$ $\{011,110,101\} \}$ corresponding the messages $0-4$ of first page.

The rows correspond to the symbol value of the second page. The sets of the codes are $\textsf{B} =\{\{000,111\}, \{001,110\}, \{010,101\},\{100,011\} \}$ corresponding the messages $0-3$ of second page. The code rate of two sets are 0.773 and 0.667, respectively. The sum rate is $R_{\text{sum}} = 1.44.$
\begin{table}[htb]
\begin{center}
 \caption{$(n=3;M_1=5,M_2=4)$-P-RIO Code} \label{TPRIO3}
  \begin{tabular}{|c||c|c|c|c|c|}
  \hline
  & $0$ & $1$ &$2$&$3$&$4$\\      \hline
$0$&000&112&121&211&122     \\ \hline
$1$&001&002&120&210&220     \\ \hline
$2$&010&102&020&201&202     \\ \hline
$3$&100&012&021&200&022     \\\hline
 \end{tabular}
 \end{center}
\end{table}
\end{example}

When the message of the first page is $1$, the message of the second page is $2$, based on the Table~\ref{TPRIO3}, we have that the cell state is $102$.

Threshold level between level $0$ and $1$ is 1, threshold level between level $1$ and $2$ is 2.
$d_{1}(102) = 101$, $d_{2}(102) = 001$.
Since $101 \in B_2$, $001 \in A_1$, thus, the message of the first page is $1$, the message of the second page is $2$.
{\myQED}

\section{Construction of $(n; M_1,M_2)$-P-RIO Code}
In this section, we give a construction of $(n; M_1,M_2)$-P-RIO code.

\subsection{Construction of $\textsf{B}$}
Let binary $n$-vector $\bm{b}_{j} = (0, b_1,b_2,\dots,b_{n-1})$ with its index $j= \sum_{l=0}^{n-2} 2^{n-1-l}b_{l}$, a decimal representaion of $\bm{b}_{j}$. We define
\begin{eqnarray}\label{Bj}
B_{j} = \{\bm{b}_{j}, \bm{\bar{b}} _{j} = \bm{1}\oplus\bm{b}_{j}\}, j=0,1, \dots, 2^{n-1}-1
\end{eqnarray}
where $\bm{1}^{n}$ is all-1 $n$-vector and notation $\oplus$ represents a modulo-2 sum of two binary vectors. Obviously, $B_{j}\cap B_{j'} = \O$ for $j \not= j'$. Thus we obtain $\textsf{B} = \{B_{0}, B_{1}, \dots, B_{M_2-1}\}$ with $M_2=2^{n-1}$

Let $w(\bm{b})$ be the Hamming weight of vector $\bm{b}$. From (\ref{Bj}) we have
\begin{equation}{\label{eq:full wright}}
w(\bm{b}_j)+w(\bm{\bar{b}}_{j})=n.
\end{equation}
Note that there always exists one vector in $B_{j}$ such that its weight is greater than or equal to $\lceil \frac{n}{2} \rceil$, where $\lceil p \rceil$ is the smallest integer greater
than or equal to $p$.


\subsection{Construction of $\textsf{A}$}


In this section, for a given $0 \leq u \leq \lceil \frac{n}{2} \rceil$,
we will show that a constant-weight set, consisting of all weight-$u$ vectors
with their non-zero elements restricted to $(2u-1)$ same positions, is
included in set $\textsf{B}$. The disjoint ones, among the constant-
weight sets with all possible permutations of  $(2u-1)$ non-zero
positions on $n$ positions for all $u=0,1, \dots, \lceil \frac{n}{2} \rceil$, forms set $\textsf{A}$.

For a given positive integer $n$, let $u$ be an integer with $0 \leq u \leq \lceil \frac{n}{2} \rceil$ and $m=2u-1$. Let $A^{(u)}_{\text{core}}$ be a collection of all $m$-vectors with Hamming weight $u$. We arrange all the $t_c$= $m \choose u$ vectors in set $A^{(u)}_{\text{core}}$ into $t_c \times m$ matrix $T^{(u)}_{\text{core}}$, called a core matrix.


From $T^{(u)}_{\text{core}}$, we give a $t_c \times n$ matrix
\begin{equation} {\label{eq:Tmatrix}}
T^{(u,n)} = [\bm{t}^{\rm T}_{0}, \bm{t}^{\rm T}_1, \dots, \bm{t}^{\rm T}_{m-1}, (\bm{0}^{t_c})^{\tt T}, \dots, (\bm{0}^{t_c})^{\tt T}]
\end{equation}
where the most left $m$ columns forms $T^{(u)}_{\text{core}}$, and the remaining $n-m$ columns $\bm{t}_{i}^{\tt T}$, $m \leq i \leq n-1$, are all-0 column vectors. Denote by $A^{(u,n)}$ a set of rows in matrix $T^{(u,n)}$.

\begin{example} \label{u2core}
For $u=2$, $m=3$, we have
$T^{(2)}_{\text{core}} =\left[\begin{array}{c }
1 1 0 \\
0 1 1 \\
1 0 1
\end{array}\right]$,
$T^{(2,4)} = \Biggl[ \begin{array}{cccc }
1 &1&0&0\\
0 &1&1&0\\
1 &0&1&0
\end{array} \Biggl], $ and $A^{(2,4)}=\{ 1100, 0110, 1010\}$.
\myQED
\end{example}

\begin{lemma}\label{lemma:Aun}
For $0 \leq u \leq \lceil \frac{n}{2} \rceil$, it follows that $A^{(u,n)} \leq \textsf{B}$.
\end{lemma}
Proof: For any $\bm{a} \in A^{(u,n)}$, we rewrite $\bm{a}=(\bm{a}(0\!:\!m\!-\!1),\bm{0}^{n-m})$ where $\bm{a}(0\!:\!m\!-\!1) \triequ (a_0,\dots, a_{m-1}) \in A^{(u)}_{\text{core}} $. From (\ref{Bj}), for any $j$, it always satisfies that
$$w(\bm{b}_{j}(0\!:\!m\!-\!1))+w(\bm{\bar{b}}_{j}(0\!:\!m\!-\!1)) = m=2u-1.$$ This implies that $w(\bm{b}_{j}(0\!:\!m\!-\!1)) \geq u$ or $w(\bm{\bar{b}}_{j}(0\!:\!m\!-\!1)) \geq u.$

Since $A^{(u)}_{\text{core}}$ includes all the $m$-vectors with weight $u$, there exists at least one, for example, $\bm{a}_{\ell_0}(0\!:\!m\!-\!1) \in A^{(u)}_{\text{core}}$, such that $\bm{a}_{\ell_0} (0\!:\!m\!-\!1) \leq \bm{b}_{j}(0\!:\!m\!-\!1)$ or $\bm{a}_{\ell_0} (0\!:\!m\!-\!1) \leq \bm{\bar{b}}_{j}(0\!:\!m\!-\!1)$. Therefore there exists at least one vector, $\bm{a}_{\ell_0} \in A^{(u,n)}$, such that $\bm{a}_{\ell_0} \leq \bm{b}_{j}$ or $\bm{a}_{\ell_0} \leq \bm{\bar{b}}_{j}$, for all $j=1,2,\dots,M_2$. This completes the proof.
\myQED

Let sequence $\Pi =(\pi_0, \pi_1, \dots, \pi_{n-1}) $ is a permutation of the integers 0 to $n-1$. Denote a column permutation on $T^{(u,n)}$ by
\begin{eqnarray}{\label{eq:T_permutation}}
T_{\Pi}^{(u,n)}=  [\bm{t}^{\rm T}_{\pi_0}, \bm{t}^{\rm T}_{\pi_1}, \dots, \bm{t}^{\rm T}_{\pi_{n-1}}].
\end{eqnarray}
Let $A_{\Pi}^{(u,n)}$ be a set of rows in matrix $T_{\Pi}^{(u,n)}$. Similar to {\sl Lemma~\ref{lemma:Aun}}, we have the following lemma.



\begin{lemma}
For $0 \leq u \leq \lceil \frac{n}{2} \rceil$ and permutation $\Pi$, it follows that $A_{\Pi}^{(u,n)} \leq \textsf{B}$.
\myQED
\end{lemma}

Since there are $(n-m)$'s all-zero columns in matrix $T_{\Pi}^{(u,n)}$, for all possible permutation, we have matrices $T_{\Pi_j}^{(u,n)}$ and their corresponding sets $A_{\Pi_j}^{(u,n)}$, $j=0,1,\dots, \binom{n}{m}-1$. Let $ \Pi_{{\ell_0}:{\ell_{m-1}}}=  \{ \pi_{\ell_0}, \pi_{\ell_1}, \dots, \pi_{\ell_{m-1}}\}$ be a set of non-zero columns' indexes in matrix $T_{\Pi}^{(u,n)}$. Denote by $| \cdot |$ a cardinality of set. The following lemma shows sufficient and necessary conditions for two permutation patterns such that their corresponding sets are disjoint.

\begin{lemma} \label{lemma:A2u}
Two sets $A^{(u,n)}_{\Pi} \cap A^{(u,n)}_{\Pi^\prime} = \O$, if and only if
\begin{equation}\label{eq:Disjoint Permutation}
 |\Pi_{{\ell_0}:{\ell_{m-1}}} \cap \Pi^\prime_{{\ell_0}:{\ell_{m-1}}}| \leq u-1.
\end{equation}
\end{lemma}

Proof: (Sufficient condition:) Assume that $| \Pi_{{\ell_0}:{\ell_{m-1}}} \cap \Pi^\prime_{{\ell_0}:{\ell_{m-1}}}| \leq u-1$. This means $ \Pi_{{\ell_0}:{\ell_{m-1}}}$ and $ \Pi^\prime_{{\ell_0}:{\ell_{m-1}}}$ have at most $u-1$ same elements. Since every $\bm{a} \in A^{(u,n)}_{\Pi}$ and every $\bm{a}^\prime \in A^{(u,n)}_{\Pi^\prime}$ have weight of $u$, it follows  $\bm{a} \neq \bm{a}^\prime$. Thus $A^{(u,n)}_{\Pi} \cap A^{(u,n)}_{\Pi^\prime} = \O$.

 \noindent (Necessary condition:) Assume that $A^{(u,n)}_{\Pi} \cap A^{(u,n)}_{\Pi^\prime} = \O$. Since weights of any vectors in these two sets are $u$, there exist  at least $u$ different elements between $ \Pi_{{\ell_0}:{\ell_{m-1}}}$ and $ \Pi^\prime_{{\ell_0}:{\ell_{m-1}}}$. Thus $| \Pi_{{\ell_0}:{\ell_{m-1}}} \cap \Pi^\prime_{{\ell_0}:{\ell_{m-1}}}| \leq u-1$. This completes the proof. \myQED

{\sl Lemma~\ref{lemma:A2u}} means that among all possible $\binom{n}{m}$ permutations $\Pi$, the permutation patterns satisfying (\ref{eq:Disjoint Permutation}) gives disjoint sets. In practical, finding these permutations is not easy. Fortunately, we can find a solution in the related research field on constant-weight codes \cite{GrahamSloane80}.  For a given codeword in a  constant-weight code with code length $n$, minimum Hamming distance $2u$, and weight $m=2u-1$, the non-zero positions in the codeword are the non-zero columns' indexes in matrix $T_{\Pi}^{(u,n)}$. Let ${M^{(u,n)}}$ be the number of constant-weight codewords. The ${M^{(u,n)}}$ constant-weight codewords give ${M^{(u,n)}}$ permutation patterns and thus provide $\{A_0^{(u,n)}, A_1^{(u,n)}, A_{M^{(u,n)}-1}^{(u,n)}\}$ with $A^{(u,n)}_{k} \cap  A^{(u,n)}_{k^\prime} = \O, k\not=k^\prime$. Here we use the same notation for $A_{\Pi_j}^{(u,n)}$ and $A_k^{(u,n)}$ for convenience.

The permutation patterns satisfying (\ref{eq:Disjoint Permutation}) and the number of ${M^{(u,n)}}$ are given in Table~\ref{Mun} for $n \leq 15$~\cite{GrahamSloane80,GrahamSloane90,Erik2000}. Note that $M^{(0,n)}=1$ for $A^{(0,n)}_{0} = \{\bm{0}^{n}\}$.

 Tables~\ref{T2n} and \ref{T3n} give all possible permutation patterns of $\Pi_{{\ell_0}:{\ell_{m-1}}}$ for forming disjoint sets.

\begin{table}[htb]
\begin{center}
        \caption{ The number of $M^{(u,n)}$ }\label{Mun}
\begin{tabular}{|c|p{0.36cm}<{\centering}|p{0.36cm}<{\centering}|p{0.36cm}<{\centering}|p{0.36cm}<{\centering}|p{0.36cm}<{\centering}|p{0.36cm}<{\centering}|p{0.36cm}<{\centering}|}
\hline
\backslashbox{$n$}{$u$} & 1     &  2     &  3   &  4                & 5           & 6  &7\\ \hline
 4 &  4                  &1                    &--             & --             &--       &--&-- \\ \hline
 5 &  5                  &2                    &1                  &   --           & --     & --&--\\ \hline
 6&  6                  &4                     &1                  &   --           & --     & --&-- \\ \hline
 7&  7                  &7                     &1                  &   1               & --     & -- &--\\ \hline
 8&  8                  &8                     &2                  &    1              & --     & -- &-- \\ \hline
 9&  9                  &12                   &3                  &    1                  &1      & --  &--\\ \hline
10& 10               &13                   &6                   &    1                  &1      & -- &--\\ \hline
11& 11               &17                   &11                  &   2                  &1      &1       &--\\ \hline
12& 12               &20                   &12                  &   3                  &1      & 1     &--\\ \hline
13& 13               &26                   &18                  &   4                  &1      & 1    &1  \\ \hline
14& 14               &28                   &28                  &   8                  &2      & 1  &1    \\ \hline
15& 15               &35                   &42                  &   15                &3      & 1  &1    \\ \hline
 \end{tabular}
\end{center}
\end{table}

\begin{table}[htb]
   \begin{center}
        \caption{ All possible patterns of $\Pi_{{\ell_0}:{\ell_{m-1}}}$ forming disjoint sets ($u=2$) }\label{T2n}
   \begin{tabular}{|c|c|l|}
                                                                       \hline & & \\ [-0.2em]
$n$  &$M^{(2,n)}$ &   $\Pi_{{\ell_0}:{\ell_{m-1}}} (u=2)$    \\ [2pt] \hline & & \\ [-0.2em]
5  & 2   & $\{0,2,3\}, \{1,3,4\}$                                  \\ [3pt] \hline & & \\ [-0.2em]
6  & 4   &  $ \{0,1,2\},\{0,3,4\},\{1,3,5\},\{2,4,5\}$         \\ [2pt] \hline & & \\ [-0.2em]
7  & 7   &  $\{0,1,2\},\{0,3,4\},\{0,5,6\},$                     \\  [2pt]
     &      &  $\{1,3,5\},\{1,4,6\},\{2,3,6\},\{2,4,5\}$ \\ [2pt] \hline && \\ [-0.2em]
8   & 8   &  $\{0,1,2\},\{5,6,7\},\{0,3,5\},\{1,3,6\},$                     \\  [2pt]
     &      &  $\{2,3,7\},\{0,4,6\},\{1,4,7\},\{2,4,5\}$ \\ [2pt] \hline && \\ [-0.2em]
9   & 12   & $\{0,1,2\},\{3,4,5\},\{6,7,8\},\{0,3,6\},$                     \\  [2pt]
     &      &  $\{1,4,7\},\{2,5,8\},\{0,4,8\},\{1,5,6\},$                     \\ [2pt]
     &      &  $\{2,3,7\},\{0,5,7\},\{1,3,8\},\{2,4,6\}$                       \\ [2pt]\hline && \\ [-0.2em]
15                   & 35   & $\{3,7,11\},\{3,8,12\},\{3,9,13\},\{3,10,14\},$ \\ [2pt]
                       &     & $\{4,7,12\},\{4,8,13\},\{4,9,14\},\{4,10,11\},$  \\[2pt]
                       &     & $\{5,7,13\},\{5,8,14\},\{5,9,11\},\{5,10,12\},$\\[2pt]
                       &     & $\{6,7,14\},\{6,8,11\},\{6,9,12\},\{6,10,13\}$\\[2pt] \hline
\end{tabular}
\end{center}
\end{table}

\begin{table}[htb]
   \begin{center}
        \caption{All possible patterns of $\Pi_{{\ell_0}:{\ell_{m-1}}}$ forming disjoint sets ($u=3$) }\label{T3n}
   \begin{tabular}{|c|c|l|}
                                                                       \hline & & \\ [-0.2em]
$n$  &$M^{(3,n)}$ &      $\Pi_{{\ell_0}:{\ell_{m-1}}}(u=3)$                   \\ [3pt] \hline & & \\ [-0.2em]                                                  7   & 1   &  $ \{0,1,2,3,4\}$                                         \\ [3pt] \hline & & \\ [-0.2em]
8   & 2   & $\{0,1,2,3,4\},\{0,1,5,6,7\}$                         \\ [3pt]  \hline && \\ [-0.2em]
9   & 3   & $\{0,1,2,3,4\},\{3,4,5,6,7\},\{0,1,6,7,8\}$       \\ [3pt]  \hline && \\ [-0.2em]
10 &6    & $\{0,1,2,3,4\}, \{0,1,5,6,7\}, \{1,2,5,8,9\},$  \\ [2pt]
    &       & $\{2,3,6,7,8\},\{3,4,5,6,9\},\{0,4,7,8,9\}$  \\ [3pt]  \hline
\end{tabular}
\end{center}
\end{table}






Moreover, for two district $u$ and $u^\prime$,  it follows that $A^{(u,n)}_{k} \cap A^{(u^\prime,n)}_{k^\prime} = \O$ since the weights in these two sets are distinct. Let
\begin{equation}{\label{eq:A}}
{\textsf A} = \{ A^{(u,n)}_{k}|  k\!=\!0,1,\dots, M^{(u,n)}\!-\!1, u\!=\!0, 1, \dots, \lceil \frac{n}{2} \rceil \}.
\end{equation}
We have the main result of this work.

\begin{theorem} \label{theorem:code}
Sets ${\textsf A}$ in (\ref{eq:A}) and ${\textsf B}$ in (\ref{Bj}) form an $$(n; M_1, M_2=2^{n-1}) \text{P-RIO code},$$ where $M_1 = \sum\limits_{i=0}^{\lceil \frac{n}{2} \rceil}M^{(i,n)}$. \myQED
\end{theorem}

Let us look more closely at $A^{(u=\lceil \frac{n}{2} \rceil,n)}_{k}$, $k\!=\!0,1,\dots, \allowbreak M^{(u,n)}\!-\!1$ in (\ref{eq:A}). In the case of odd $n$, there does not exists any all-zero column in $ T^{( \lceil \frac{n}{2} \rceil,n)}$ of (\ref{eq:Tmatrix}). Thus sets $A_{\Pi}^{( \lceil \frac{n}{2} \rceil,n)}$ from any possible permutations are the same as $ A^{( \lceil \frac{n}{2} \rceil,n)}$ itself, and thus $M^{(\lceil \frac{n}{2} \rceil,n)} =1$. Furthermore, we observe that  $A^{( \lceil \frac{n}{2} \rceil,n)}=E^{(\lceil \frac{n}{2} \rceil,n)}$, where $E^{(u,n)}$ is the set of all $n$-vector with Hamming weight $u$.

In the case of even $n$, there exists one all-zero column in $ T^{( \frac{n}{2},n)}$ of (\ref{eq:Tmatrix}), and thus $n$ possible permutation patterns gives $n$ possible matrices, $T_{\Pi}^{( \frac{n}{2},n)}$s. According to ${\sl Lemma~\ref{lemma:A2u}}$, the corresponding $n$ possible sets, $A_{\Pi}^{( \frac{n}{2},n)}$s, are joint each other. As a result, we choose one, e.g., $A^{( \frac{n}{2},n)}$, among $A_{\Pi}^{( \frac{n}{2},n)}$s, in our proposed code in {\sl Theorem~\ref{theorem:code}}, and thus $M^{(\frac{n}{2},n)} =1$. Different from the case of odd $n$, we observe that $A^{(\frac{n}{2}, n)}  \subset E^{(\frac{n}{2},n)}$ since $A^{(\frac{n}{2}, n)}$ consists of only the weigh-$\frac{n}{2}$ vectors with the most right bit being 0. This observation motives us to improve our proposed codes in {\sl Theorem~\ref{theorem:code} } by adding a supplemental set to ${\textsf A}$.

Specifically, given an even $n$, let $\bar{A}^{( \frac{n}{2},n )}=\{ \bm{1} \oplus \bm{a} | \bm{a} \in {A}^{( \frac{n}{2},n)} \}$ and  $A_{\text{sup} }=\bar{A}^{( \frac{n}{2},n )} \cup E^{(\frac{n}{2}+1,n)}$. Adding $A_{\text{sup} }$ to ${\textsf A}$ of (\ref{eq:A}) provides ${\textsf A}^\prime=\{ {\textsf A}, A_{\text{sup} }\}$.

\begin{theorem}\label{theorem:even}
For a given even $n$, sets ${\textsf A^\prime}$ and ${\textsf B}$ in (\ref{Bj}) form an $$(n; M_1^\prime, M_2=2^{n-1})~\text{P-RIO code},$$ where $M_1^\prime=M_1+1$.
\end{theorem}
Proof: We first show that $A_{\text{sup} } \leq {\textsf B}$. We partition ${\textsf B}$ into two parts ${\textsf B}^{\frac{n}{2}}$ and ${\bar{\textsf B}}^{\frac{n}{2}}$. The first part ${\textsf B}^{\frac{n}{2}}$ consists of $B_j$s whose elements are with weight $n/2$. The remaining $B_j$s form ${\bar{\textsf B}}^{\frac{n}{2}}$. Since $w(\bm{b}_j) + w(\bar{\bm{b}}_j) =n$, every $B_j \in {\bar{\textsf B}}^{\frac{n}{2}}$ has a vector whose weight is greater than $n/2+1$. Therefore we have $E^{(\frac{n}{2}+1,n)} \leq {\bar{\textsf B}}^{\frac{n}{2}}$.

We now show $ \bar{A}^{( \frac{n}{2},n )} < {\textsf B}^{\frac{n}{2}}$. Among $\bar{A}^{( \frac{n}{2},n )}$, the most right bit of every vector is 1, and the remaining bits forms the all subvectors with weight $n/2-1$. In ${\textsf B}^{\frac{n}{2}}$, one of two vectors always has the most right bit being 1 and remaining bits be weight $n/2-1$. Therefore $\bar{A}^{( \frac{n}{2},n )} \leq {\textsf B}^{\frac{n}{2}}$. It follows that $A_{\text{sup} } \leq {\textsf B}$.

It remains to show $A_{\text{sup} } \cap {\textsf A} =\O$. Since the most right bit in $\bar{A}^{( \frac{n}{2},n )}$ is 1 while that in $A^{( \frac{n}{2},n)}$ is 0, it follows that $ \bar{A}^{( \frac{n}{2},n )}  \cap {A}^{( \frac{n}{2},n)} = \O$ and thus $\bar{A}^{( \frac{n}{2},n )} \cap {\textsf A}= \O$. Also we have $E^{(\frac{n}{2}+1,n)} \cap {\textsf A}= \O$ because weights in the two sets are distinct. Therefore $A_{\text{sup} } \cap {\textsf A} =\O$. This completes the proof.
\myQED

\subsection{Code Rate of Two-Write Unrestricted-Rate  P-RIO-Code}

Based on $M^{(u,n)}$ in Table~\ref{Mun}, we have $R_{\text{sum}}$ of $(n;M_1,M_2)$ P-RIO codes in Table~\ref{coderate}, compared with that of $(n;M,M)$ fixed-rate P-RIO code \cite{Yaakobi2016}. From the table, we see that the sum rates of our P-RIO codes are higher than those of the fixed P-RIO codes given in \cite{Yaakobi2016} when $n=3,4,5,6$. When $n >6$, there is no data in \cite{Yaakobi2016} since construction complexity is very high.

\begin{table}[htb]
\begin{center}
  \caption{$(n;M_1,M_2;2)$ P-RIO codes for $3 \leq n \leq 15$}\label{coderate}
  \begin{tabular}{|c|c|c|c|c|c|}
  \hline
$n$ & $M_1 $  & $M_2 $  & $R_{\text{sum}}$  &$M $ \cite{Yaakobi2016} & $R'_{\text{sum}}$\cite{Yaakobi2016}     \\ \hline
3 & 5    &4        & 1.44        &4 &1.333              \\ \hline
4 & 7    &8        & 1.452        &7 &1.4037           \\ \hline
5 & 9    &16       & 1.434        &11 &1.384              \\ \hline
6 & 13    &32        & 1.45        &19 &1.41              \\ \hline
7 &  17   &64      & 1.44        &  /&/ \\ \hline
8& 21    &128      & 1.424        &/ &/ \\ \hline
9& 27    &256      & 1.417        &/ &/ \\ \hline
10& 33    &512      & 1.404        &/ &/ \\ \hline
11& 44    &1024      & 1.406        &/ &/ \\ \hline
12& 51    &2048      & 1.389        &/ &/ \\ \hline
13& 65    &4096      & 1.386        &/ &/ \\ \hline
14& 84    &8192      & 1.385        &/ &/ \\ \hline
15& 114    &16384     & 1.389        &/ &/ \\ \hline
\end{tabular}
\end{center}
\end{table}

\subsection{Examples}
\begin{table*}[htb]
   \begin{center}
     \caption{(5;9,16)-P-RIO Code}\label{TPRIO5}
   \begin{tabular}{|c||c|c|c|c|c|c|c|c|c|}
  \hline
   & $0$ & $1$ &$2$&$3$&$4$ &$5$&$6$&$7$ &$8$    \\ \hline
$0$&00000&21111&12111 &11112&11121&11211&21211& 12121&22211                            \\ \hline
$1$&00001&21110&12110 &00002&11120&11210&21210&12120& 22210                        \\ \hline
$2$&00010&21101&12101 &11102&00020&11201&21201&12102&22201                       \\ \hline
$3$&00011&21100&12100& 00012&00021&11200& 21200&00022&22200                            \\ \hline
$4$&00100&21011&12011&11012&11021&00200& 21021&12012&22021                      \\ \hline
5    &00101&21010&12010&00102&11020&00201&21020&12020&22020                        \\ \hline
6    &00110&21001&12001& 11002 &00120&00210&00220&12002&22002                            \\ \hline
7    &00111&21000&12000&00112&00121&00211&00221&00122&00222                           \\ \hline
8    &01000&20111&02000 &10112&10121&10211&20121&10122&10222                            \\ \hline
9    &01001&20110&02001&01002&10120&10210&20120&02002&20220                          \\ \hline
10&01010&20101& 02010&10102&01020&10201&20201&02020&20202                         \\ \hline
11&01011&20100&02011&01012&01021&10200&20200&02012&02022                          \\ \hline
12&01100&20011&02100& 10012 &10021&01200&20021&10022&20022                            \\ \hline
13&01101&20010&02101&01102&10020&01201&20020&02102&02202                           \\ \hline
14&01110&20001&02110&10002&01120&01210&01220&02120&02220                          \\ \hline
15&01111&20000&02111&01112 &01121&01211&01221&02112&01222                           \\ \hline
\end{tabular}
\end{center}
\end{table*}

Now we give two examples of P-RIO codes with code length of $n=4,5$.

\begin{example}
For $n=4$, by (\ref{Bj}), we have the set of second page of
\begin{eqnarray*}
\textsf{B}&=&  \{\{0000,1111\}, \{0001,1110\}, \{0010,1101\},\\
&&~ \{0011,1100\},\{0100,1011\}, \{0101,1010\}, \\
&&~  \{0110,1001\},\{0111,1000\}\},
\end{eqnarray*}

Next, we construct $\textsf{A}'$ as follows.
When $u=0$, $T^{(1)}_{\text{core}} = [0]$, and then $A^{(0,4)}_{0}= \{0000\}$.
When $u=1$, $T^{(1)}_{\text{core}} = [1]$, we have that $A^{(1,4)}_{0} = \{1000\}, A^{(1,4)}_{1} = \{0100\}, A^{(1,4)}_{2}= \{0010\}, A^{(1,4)}_{3} = \{0001\}$.
When $u=2$, from Example \ref{u2core} we have $A^{(2,4)}_{0}=\{1100,0110,1010\}$.

Since $n$ is even, $\bar{A}^{( 2,4 )}= \{0011,1001,0101\}$ and $ E^{(3,4)} = \{1110, 1101,1011,0111\}$, from Theorem \ref{theorem:even}, we have $A_{\text{sup}}=\{0011,1001,0101,1110,1110, 1101,1011,0111\}$.

Therefore, we have $\textsf{A}' = \{A^{(0,4)}_{0}, A^{(1,4)}_{0}, A^{(1,4)}_{1}, A^{(1,4)}_{2}, \allowbreak A^{(1,4)}_{3}, \allowbreak  A^{(2,4)}_{0}, A_{\text{sup}} \}$.
From Theorem~\ref{theorem:even}, $\textsf{A}'$ and $\textsf{B}$ provide $(4; 7,8)$-P-RIO code as shown in Table.~\ref{TPRIO4}.
\myQED
\end{example}

\begin{table}[htb]
\begin{center}
  \caption{(4;7,8)-P-RIO Code}\label{TPRIO4}
  \begin{tabular}{|c||c|c|c|c|c|c|c|}
  \hline
   & $0$ & $1$ &$2$&$3$&$4$ &$5$&$6$                                    \\ \hline
$0$&0000&2111&1211&1121&1112&2200&1122 \\ \hline
$1$&0001&2110&1210&1120&0002&2210&2220 \\ \hline
$2$&0010&2101&1201&0021&1102&2201&2202 \\ \hline
$3$&0011&2100&1200&0021&0012&2200&0022 \\ \hline
$4$&0100&2011&0200&1021&1012&2021&2022 \\ \hline
$5$&0101&2010&0201&1020&0102&2020&0202 \\ \hline
$6$&0110&2001&0210&0120&1002&0220&2002 \\ \hline
$7$&1000&2000&0211&0121&0112&0221&0222 \\ \hline
  \end{tabular}
  \end{center}
\end{table}

\begin{example} For $n=5$, from (\ref{Bj}), we have
\begin{eqnarray*}
\textsf{B}&=&  \{\{00000,11111\}, \{00001,11110\}, \{00010,11101\},\\
&&~ \{00011,11100\},\{00100,11011\}, \{00101,11010\}, \\
&&~  \{00110,111001\},\{00111,11000\}, \{01000, 10111\}, \\
&&~  \{01001,10110\}, \{01010,10101\}, \{01011,10100\}, \\
&&~ \{01100,10011\}, \{01101,10010\}, \{01110,10001\}, \\
&&~ \{01111,10000\}\}
\end{eqnarray*}

When $u=0$ and $u=1$, we have $A^{(0,5)}_{0}=\{00000\}$, and $A^{(1,5)}_{0}=\{00001\}$, $A^{(1,5)}_{1}=\{00010\}$, $A^{(1,5)}_{2}=\{00100\}$, $A^{(1,5)}_{3}=\{01000\}$, $A^{(1,5)}_{4}=\{10000\}$.

When $u=2$, from Table~\ref{T2n}, we have
$A^{(2,5)}_{0} = \{10010,10100,10100\}$ and $A^{(2,5)}_{1}= \{01010,01001,00011\}$.

When $u=3$, we have $A^{(3,5)}_{0}= \{00111,01110,11100,$ $01011,10110,01101,11010,10011,11001,10101\}$.

Finally, we obtain the (5;9,16)-P-RIO code in Table.~\ref{TPRIO5}.
\end{example}

\section{Conclusion}
In this paper, we proposed a coding scheme for two-page unrestricted-rate P-RIO code that each page may have different code rates. Our coding scheme is constructive, and the code length is arbitrary. The sum rates of our proposed codes are higher than those of conventional fixed-rate P-RIO codes in \cite{Yaakobi2016}.

\section*{Acknowledgement}

This work was supported by the Japan Society for the Promotion of Science through the Grant-in-Aid for Scientific Research (C) under Grant 16K06373, in part by MEXT through the Strategic Research Foundation at Private Universities (2014-2018) under Grant S1411030, and the Advanced Storage Research Consortium and JSPS KAKENHI Grant 15K00010.

\end{document}